\documentstyle[prb,aps,multicol,epsfig]{revtex}
\tighten

\begin{document}
\draft
\title{Density of States and Conductivity of Granular Metal or Array of Quantum Dots}

\author{Jingshan Zhang and Boris I. Shklovskii}

\address{Theoretical Physics Institute, University of Minnesota, 116 Church
  Street S.E., Minneapolis, Minnesota 55455}

\date{\today}
\maketitle

\begin{abstract}
The conductivity of a granular metal or an array of quantum dots
usually has the temperature dependence associated with variable
range hopping within the soft Coulomb gap of density of states.
This is difficult to explain because neutral dots have a hard
charging gap at the Fermi level. We show that uncontrolled or
intentional doping of the insulator around dots by donors leads to
random charging of dots and finite bare density of states at the
Fermi level. Then Coulomb interactions between electrons of
distant dots results in the a soft Coulomb gap. We show that in a
sparse array of dots the bare density of states oscillates as a
function of concentration of donors and causes periodic changes in
the temperature dependence of conductivity. In a dense array of
dots the bare density of states is totally smeared if there are
several donors per dot in the insulator.

\end{abstract}

\begin{multicols}{2}

\section{Introduction}

Conduction of samples where metallic granules are surrounded by
some kind of insulator (granular metals) have been studied
intensively for
decades~\cite{sheng,experiments,crossover,book,chuiandgd,Pollak,chui,Cuevas,Baskin}.
If volume fraction of the metal $x$ is large, metallic granules
touch each other and conductivity is metallic. When $x$ decreases
and crosses percolation threshold $x_c$, granules become isolated
from each other and granular metal goes through metal-insulator
transition. It is generally
observed~\cite{sheng,experiments,crossover} that at the insulator
side of transition the temperature dependence of conductivity
obeys
\begin{equation} \label{a1}
\sigma = \sigma_{0} \exp\left[-\left(\frac{T_{0}}{T}\right)^{1/2}
\right].
\end{equation}
Recently, similar temperature dependence was observed in doped
systems of self-assembled germanium quantum dots on the silicon
surface~\cite{Yakimov} and CdSe nanocrystal thin
films~\cite{Guyot}. Below we talk about both granular metal and
semiconductor dot structures universally using the word dot for
brevity.

\begin{figure}[ht]
\begin{center}
\includegraphics[height=0.15\textheight]{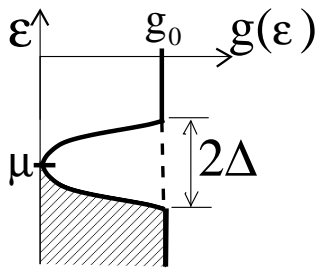}
\end{center}
\caption{The shape of Coulomb gap in the vicinity of the Fermi
level. Bare density of states in the absence of long range Coulomb
interaction is shown by the dashed line. Occupied states are
shaded. } \label{fig:gap}
\end{figure}

In doped semiconductors the temperature dependence of
Eq.~(\ref{a1}) is also widely observed at low
temperatures~\cite{book}. It is interpreted as the variable range
hopping conductivity between impurities in the presence of the
Coulomb gap~\cite{book,coulombgap} of the density of states (DOS)
and is called the Efros-Shklovskii (ES) law. In a $n$-type lightly
doped compensated semiconductor all acceptors are negatively
charged and randomly situated, an equal number of donors are
charged positively. Together all random charges create a random
potential shifting donor levels up and down. This results in
finite bare DOS $g_{0}(\mu)$ at the Fermi level $\mu$. Long range
Coulomb interaction of localized electrons then produces the
parabolic Coulomb gap in DOS
\begin{equation} \label{parabola}
g(\varepsilon)=\frac{3}{\pi} \kappa ^3 \varepsilon ^2 /e^6
\end{equation}
at the Fermi level (Fig.~\ref{fig:gap}) leading to ES law with
$T_0 = C \cdot e^2/\kappa \xi$, where $C$ is a constant factor,
$\kappa$ is dielectric constant and $\xi$ is the localization
length.

\begin{figure}[ht]
\begin{center}
\includegraphics[height=0.15\textheight]{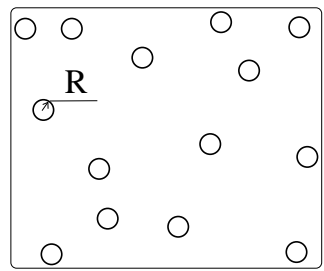}
\end{center}
\caption{A sparse array of same size neutral metallic dots
surrounded by an insulator.} \label{fig:x}
\end{figure}
\begin{figure}[ht]
\begin{center}
\includegraphics[height=0.19\textheight]{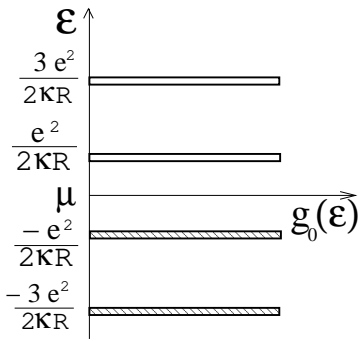}
\end{center}
\caption{The bare density of ground states (BDOGS) of a clean
system of identical dots consists of equidistant peaks. Occupied
states are shaded.} \label{fig:init}
\end{figure}

In contrary to a doped semiconductor, in an array of neutral dots
the bare DOS at the Fermi level $g_{0}(\mu) =0$ and there is no
justification for ES law, which requires a non-zero $g_{0}(\mu)$
to begin with. Let us consider a sparse array of dots with the
same radius $R$ shown on Fig.~\ref{fig:x}. Charging energy levels
of such array are shown in Fig.~\ref{fig:init}. Here we assume
that dots are large enough, so that spacing between charging
levels $e^2/\kappa R$ is much greater than spacing between quantum
levels of the dot with a given charge. Empty peaks in
Fig.~\ref{fig:init} correspond to energies necessary to charge a
neutral dot by the first, second and so on electrons transferring
them from the macroscopic piece of the same metal. Shaded (filled)
peaks correspond to, taken with the sign minus, energies necessary
to extract first, second and so on electrons from the dot, or in
other words, this is density of states of holes. The Fermi level
of the array is at zero between two peaks and coincides with the
Fermi level of macroscopic piece of the same metal.

We want to emphasize that in each dot we are dealing with the
ground state at a given number of electrons and exclude excited
states (higher quantum levels) because the ground states of dots
determine the low temperature hopping transport. Indeed, in the
Miller-Abrahams network~\cite{MA} of resistances connecting all
dots, the exponentially large activation factor of each resistance
depends on probabilities of occupation of a dot by a given number
of electrons. Exponential temperature dependencies of these
probabilities, as well as the partition functions of dots are
determined by ground state energies~\cite{criticismBaskin}. Thus,
the density of states of the array we need can be called bare
density of ground states (BDOGS).

Note that in a lightly doped semiconductor with several equivalent
conduction band minima each donor has excited states close to the
ground state, i. e. situation is similar to large dots. Still
exponential temperature dependence of hopping conductivity depends
only on BDOGS~\cite{book,MA}.

Let us illustrate the role of BDOGS in a clean sparse array
(Fig.~\ref{fig:init}). In this case, conductivity requires
activation of electron-hole pairs or in other words, transfer of
an electron between two originally neutral dots. Obviously
concentrations of positive and negative dots and the hopping
conductivity obeys
\begin{equation} \label{condu}
\sigma=\sigma_0 \exp[-E_c/2k_BT]
\end{equation}
independently of excited states, (Same result can be obtained in
Miller-Abrahams resistor network approach.) where $E_c=e^2/ \kappa
R$ is the charging energy.

How then can we explain observation of ES law? Apparently BDOGS
shown in Fig.~\ref{fig:init} should be smeared in the vicinity of
the Fermi level due to some kind of disorder.

A simplest source of disorder is distribution of sizes or
capacitances of dots. Indeed, charging energies of larger dots are
smaller and this can result in the low energy tail of the first
empty peak of BDOGS and symmetric high energy tail of the first
occupied peak. Still in a neutral system these tails do not
overlap and $g_0(\mu)$ is zero, so that this kind of disorder does
not lead to ES law. Sheng et al.~\cite{sheng} assumed that in a
reasonably dense system of neutral granules there is a special
distribution of distances between granules or their mutual
capacitances, which can lead to ES law. This assumption was found
incompatible with other experiments~\cite{chuiandgd}. But more
importantly it was noticed~\cite{Pollak} that in a system of large
granules made of the same metal all granules remain neutral in
ground state at any distribution of mutual and individual
capacitances. Therefore, inter-granular excitation of
electron-hole pairs has a gap.

This leads to an important conclusion that granular metal may have
finite BDOGS $g_0(\mu)$ and show ES conductivity only if in the
ground state of the system granules are
charged~\cite{book,chuiandgd,Pollak,chui,Cuevas,Baskin}. Formally
one can imagine that this happens if granules have different work
functions~\cite{book,Pollak}. Several possible mechanisms of such
fluctuations were discussed in literature. Chui~\cite{chui}
suggested that very small dots can charge big dots in the case
when the former are so small that their quantum level spacing
exceeds the charging energy. We concentrate on the system of large
enough dots and, therefore, ignore this possibility. Cuevas et
al.~\cite{Cuevas} claimed that even in large dots there are large
fluctuations of the Fermi level $\delta E_F$ due to random
positions of neutral impurities in different dots. To our mind,
this possibility can be rejected using for $\delta E_F$ a simple
estimate of typical fluctuation of the average potential in the
metallic dot due to fluctuation of number of impurities there. For
three dimensional case one easily gets $\delta E_F \sim E_{F}
c^{1/2}/(k_FR)^{3/2}$, where $c$ is relative concentration of
impurities in the dot, $k_{F}$ is Fermi wave-vector, and we assume
that size of an impurity is of the order of $k_F^{-1}$ and its
potential $U \sim E_F$. At large $R$ this energy is apparently
smaller than charging energy $e^2/\kappa R$ even at $c\sim 1$, so
that dots remain neutral. Baskin and Entin~\cite{Baskin}
considered fluctuations of surface part of the dot energy as a
reason of ionization of dots. Again if we assume that these
fluctuations are result of random distribution of point-like
impurities located on the dot surface we come to conclusion that
corresponding fluctuation of energy decreases like $1/R^{2}$ and
can not compete with the charging energy. Thus, simplest internal
mechanisms of fluctuations of the work function mentioned above
are too weak to charge array of three-dimensional dots.

In this paper we study models of arrays of dots affected by
external disorder, where the origin of charging and BDOGS $g_0$ is
transparent and in some cases controllable. We assume that the
insulator between clean dots has a concentration of donors $N_D$
with electron energy $E_D$ higher than Fermi energy $\mu$ in dots,
so that at low temperatures all donor donate an electron to dots
and charge them. For a given concentration of dots $N$
(Fig.~\ref{fig:x}) we can introduce the average number of extra
electrons per dot $\nu \equiv N_D /N$, which by analogy with
quantum Hall effect can be called the filling factor. Maximum
number of electrons which can be added to a dot is $n_{max}\simeq
(E_D-\mu)/(e^2/\kappa R)$. We assume that $\nu<n_{max}$ so that
all donors lose their electrons. In the semiconductor language we
are dealing with compensated $p$-type semiconductor where dots
play the role of multi-charged acceptors.

\begin{figure}[ht]
\begin{center}
\includegraphics[width=0.2\textwidth]{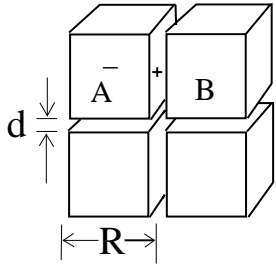}
\end{center}
\caption{ A ``super-dense" array of dots. Only one donor is shown
by + and electron donated by it is shown by -. } \label{fig:dense}
\end{figure}

In Sec. II we summarize our main results for BDOGS at the Fermi
level $g_0(\mu)$ as function of the filling factor $\nu$. In Sec.
III the sparse 3D array of dots (Fig.~\ref{fig:x}) is discussed.
In Sec. IV we extend this discussion to 2D array. In Sec. V we
study in detail the super-dense 3D array shown in
Fig.~\ref{fig:dense}, which is the other extreme, and then comment
on the realistic moderately dense array (Fig.~\ref{fig:almost}).
We also see how the properties of sparse 3D arrays cross over to
dense arrays. After studying different arrays one can see that
external doping brings about non-zero $g_0(\mu)$ and leads to ES
law.

\section{Summary of results}

\begin{figure}[ht]
\begin{center}
\includegraphics[height=0.15\textheight]{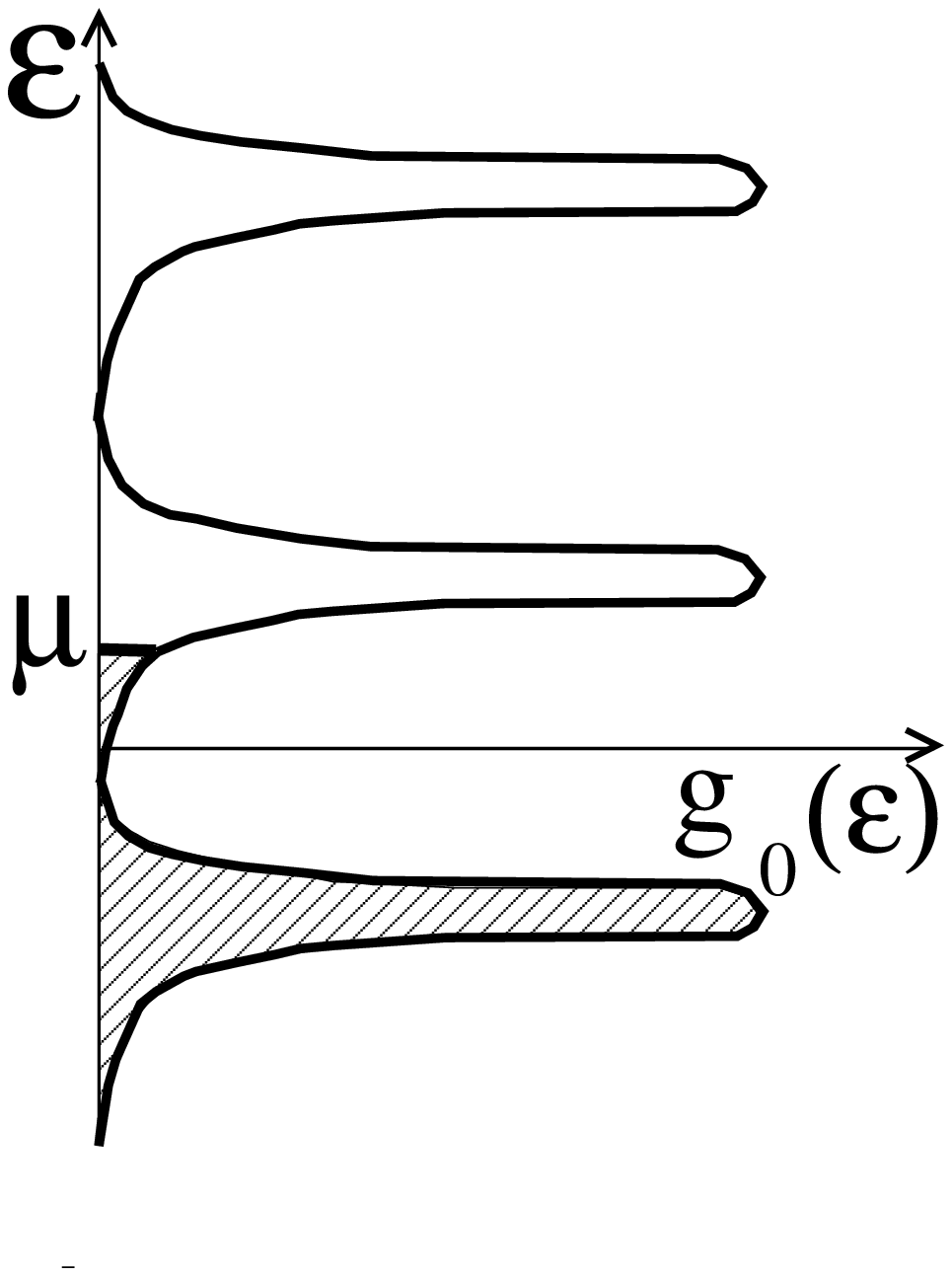}
\end{center}
\caption{BDOGS $g_{0}(\varepsilon)$ at a certain filling factor.
Occupied states are shaded.} \label{fig:low}
\end{figure}

We start from sparse three dimensional arrays of dots of the same
radius $R$, which are randomly situated in space with
concentration $N \ll (4\pi R^{3}/3)^{-1}$. We assume that the dots
are big enough so that Coulomb effect over-weights quantum level
spacing. For sparse arrays of dots, doping introduces two types of
charges: positive, empty donors and negatively charged dots. Both
are randomly situated and create random potentials growing with
$\nu$. These charges result in two effects (Fig.~\ref{fig:low}).
Firstly, the $\delta$-shaped peaks in BDOGS $g_{0}(\varepsilon)$
become somewhat smeared, since the energy it takes to bring an
electron to a dot is affected by the random potential (the effect
is similar to that of gedanken random gate potential). As a
result, each peak gets tails. Secondly, electrons coming from
donors fill some dot states and hence move the position of the
Fermi level up.

\begin{figure}[ht]
\begin{center}
\includegraphics[height=0.24\textheight]{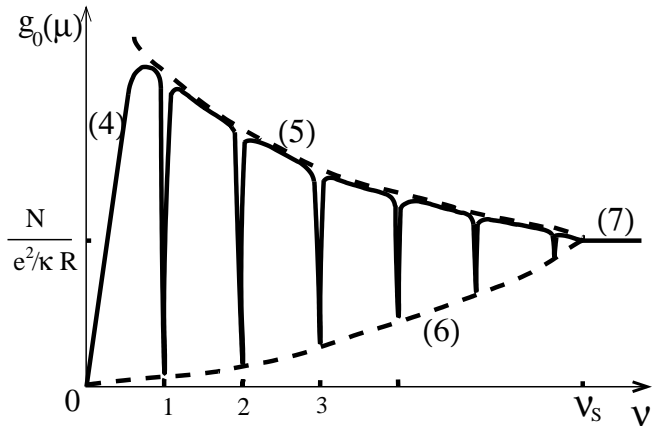}
\end{center}
\caption{BDOGS at the Fermi level $g_0(\mu)$ of a sparse 3D array
as a function of filling factor (solid line). The reference to the
equation appropriate for a part of the curve is shown next to it.
Dashed lines describe locations of minima and maxima of the
oscillating part.} \label{fig:curve}
\end{figure}

As a result $g_0(\mu)$ may oscillate with $\nu$. For example, at
$\nu =1/2$ the Fermi level $\mu$ is in the middle of the BDOGS
peak and for that reason $g_0(\mu)$ can be rather large. On the
other hand, at $\nu =1$ the Fermi level $\mu$ is in a tail between
two BDOGS peaks, and $g_0(\mu)$ tends to be much lower. Actually
the total dependence of $g_0(\mu)$ on $\nu$ is somewhat more
complicated and consists of three parts, the growing part, the
oscillating part and the saturation part (Fig.~\ref{fig:curve}).
The first (growing) part takes place at small $\nu$, where the
situation is similar to $p$-type semiconductors at low degree of
compensation~\cite{compensated}. We get BDOGS growing linearly
with $\nu$
\begin{equation} \label{bb1}
g_<(\mu)\sim\kappa \nu N^{2/3}/{e}^2~~~~~~~~~~(\nu << 1).
\end{equation}
In the second (oscillating) part, $\mu$ typically dwells in BDOGS
peaks,
\begin{equation} \label{bb2}
g_{max}(\mu) \sim\frac{N^{2/3} \kappa}{e^2
{\nu}^{4/3}}~~~~~~~~~~(1 < \nu < \nu_s).
\end{equation}
When $\nu$ is very close to integers, the Fermi level drop into
minima of $g_{0}(\varepsilon)$ where
\begin{equation} \label{bb3}
g_{min}(\mu) \sim\frac{{\nu}^{8/3} N^{5/3} R^2}{e^2/(\kappa
R)}~~~~~(1 < \nu < \nu_s).
\end{equation}
Here $\nu_s \sim1/(N R^3)^{1/4}>>1$ is the filling factor at which
oscillations become relatively small and the third (saturation)
part starts over. This happens because fluctuations of the Coulomb
potential are so big that BDOGS is almost uniform everywhere
\begin{equation} \label{bb4}
g_s \sim\frac{N}{e^2/(\kappa R)}~~~~~~~~~~~(\nu>\nu_s).
\end{equation}
The number of large oscillations $\nu_s$ is big only when dots are
far from each other. Here and below we often omit numerical
coefficients.

For less ideal arrays where the dots have slightly different
sizes, the distribution of sizes can wash away the oscillations to
a certain extent, making the line of maxima lower and the line of
minima higher. As a result the two lines approach each other
faster ($\nu_s$ is smaller) than the situation of identical sizes.

\begin{figure}[ht]
\begin{center}
\includegraphics[width=0.36\textwidth]{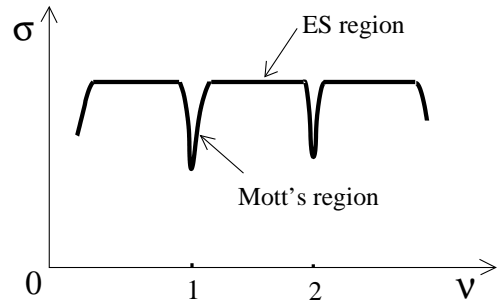}
\end{center}
\caption{Ranges of ES and Mott's laws alternating with growth of
filling factor $\nu$ at a given low temperature.} \label{fig:laws}
\end{figure}

Oscillations of $g_0(\mu)$ lead to periodic transitions between
the ES law and the Mott's law at a given low temperature as shown
in Fig.~\ref{fig:laws}. This happens because in the very vicinity
of $\mu$ the long range Coulomb interaction creates the parabolic
Coulomb gap (not shown in Fig.~\ref{fig:low}). The width of
Coulomb gap as follows from Eq.~(\ref{parabola}) and
Fig.~\ref{fig:gap} depends on $g_0$
\begin{equation} \label{width}
\Delta \sim \sqrt {g_0(\mu) e^6 / \kappa ^3   } .
\end{equation}
At large $g_0$ the Coulomb gap is wide ($\Delta >>k_BT$) and the
conductivity obeys the ES law (Eq.~(\ref{a1})), which does not
depend on $g_0$. At a very small $g_0$ the width of the Coulomb
gap is smaller than $k_BT$ and becomes irrelevant for
conductivity. In this case we arrive at the Mott's law region
\begin{equation} \label{j2}
\sigma = \sigma_0 \cdot \exp \left[- \left( \frac{\beta}{g_0(\mu)
a^3 T} \right) ^{1/4} \right],
\end{equation}
with strong dependence on $ g_0(\mu)$. Here $\beta$ is a numerical
coefficient~\cite{book}.

Oscillations of BDOGS $ g_0(\mu)$ with $\nu$ lead to oscillations
of other measurable quantities. It was shown~\cite{Baranovskii}
that in systems with the Coulomb gap BDOGS $ g_0(\mu)$ plays the
role of the thermodynamic density of states $dn/d \mu$, where $n$
is concentration of electrons, so that it determines screening
radius of the system $r_s=(4\pi g_0 e^2/\kappa)^{-1/2}$ used below
for self-consistent calculations of BDOGS. One can also measure
BDOGS at the Fermi level with help of extremely low temperature
(lower than typical quantum level spacing) specific heat and
microwave absorption~\cite{book,ESreview}

Until now we talked about 3D arrays of dots. In Sec. IV we
consider arrays of dots located in a 2D plane. In a 2D array dots
can be charged by donors located in parallel to the plane
($\delta$-layer)~\cite{Yakimov,2Ddoped}. In 2D case, however,
there is a more practical way to charge dots using a metallic gate
parallel to the plane of dots. At $\nu>>1$ results for $g_0(\mu)$
are almost independent on the way of charging, because most of the
random potential is created by dot charges. Dependence of
$g_0(\mu)$ on $\nu$ in 2D qualitatively looks similar to the 3D
case discussed above, but quantitatively it is somewhat different
(see Sec. IV).

Dependence of $g_0(\mu)$ on $\nu$ in 2D should leads to
oscillations of conductivity. Some oscillation were observed in
experiments~\cite{Yakimov,2Ddoped}. But there are additional ways
to measure this dependence in a gated in 2D structure.
First~\cite{Yakimov}, one can study change of conductivity where
the gate voltage charges dots. Second~\cite{Ef}, one can study
inverse small signal AC capacitance of unit area $1/C$, which is
related to the two-dimensional screening radius of the dot system
$r_s = \kappa/(2\pi g_0 e^2)$ by equation
\begin{equation} \label{k1}
1/C = 4 \pi \kappa (d_0 + r_s),
\end{equation}
where $d_0$ is distance from gate to the plane of dots. This
method is similar to magneto-capacitance measurements in quantum
Hall effect~\cite{Smith}.

Until now we talked about sparse arrays of dots. For a sparse
array of dots the variable range hopping conductivity is
measurable only when the tunneling length in insulator is large
enough. This probably can be achieved in some cases in narrow gap
semiconductors. Otherwise conductivity is so small that it can not
be measured. In this case, thermodynamic measurements of
oscillating quantities may be more convenient.

\begin{figure}[ht]
\begin{center}
\includegraphics[width=0.36\textwidth]{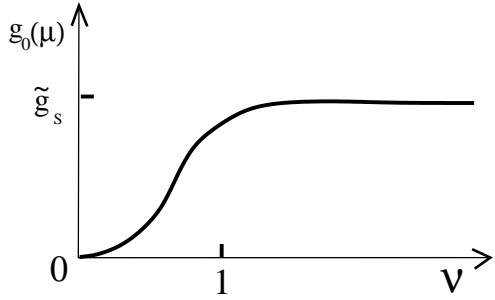}
\end{center}
\caption{BDOGS at the Fermi level $g_0(\mu)$ of a ``super-dense"
array as a function of filling factor $\nu$.} \label{fig:curvenew}
\end{figure}

It is much easier to measure conductivity in a dense array of
dots. Actually most of cited experiments are done at $NR^3 \sim
1$. In Sec. V we study clean metallic cubes separated by thin
layers of an insulator doped by donors in the ``super-dense" limit
$d<<R$ (Fig.~\ref{fig:dense}). It is assumed that the tunneling
conductance between two cubes $G<<e^2/h$. We show that in the
dense array BDOGS as a function of the average number of donors
per dot $\nu$ behaves as shown in Fig.~\ref{fig:curvenew}. At
$\nu<<1$ BDOGS is very small and grows super-linearly with $\nu$
(like $\nu^3$), and at $\nu \sim 1$ it saturates at the value
$\tilde{g}_s \sim \kappa / (e^2RDd)$.

\section{BDOGS in sparse three-dimensional arrays.}

\vspace{8pt} { \bf  1. Saturation of BDOGS} \vspace{8pt}

With enough impurities the BDOGS is strongly smeared and it
approaches an averaged value given by Eq.~(\ref{bb4}). This
happens when the fluctuation of potential $\gamma$ is large
enough, namely
\begin{equation} \label{flucnew}
\gamma > e^2/\kappa R~.
\end{equation}

In order to find the critical $\nu_s$ of the saturation BDOGS, let
us study the fluctuation of electric potential. For a given
quantum dot, all the charges within a distance of the screening
radius $r_s$ contribute into a collective Coulomb potential. There
are two kinds of charges, positive donors and negative dots. Let
us compare the charge fluctuations made by them. The average
number of impurities within $r_s$ is $N \nu r_s ^3$, with a
typical charge fluctuation $e \sqrt{N \nu r_s ^3}$. For the
charged dots, the typical fluctuation of dot number in $r_s$ is
$\sqrt{N r_s ^3}$, and each dot carries a charge of $e\nu$ on
average, so the typical charge fluctuation is $\nu e \sqrt{N r_s
^3}$. At $\nu>>1$ the charge fluctuations produced by charged dots
are larger than that of donors, so that the fluctuation of the
potential energy of an electron is mostly related to charge
fluctuations of dots
\begin{equation} \label{d1}
\gamma\sim \frac{\nu e^2}{\kappa r_s} \sqrt{N r_s ^3}~.
\end{equation}

On the other hand, the BDOGS at the Fermi level $g_0(\mu)$
determines the linear screening radius due to redistribution of
electrons between dots
\begin{equation} \label{f1}
r_s= \frac{1}{\sqrt{4 \pi
g(\mu)e^2/\kappa}}\sim\frac{1}{\sqrt{NR}}.
\end{equation}

With the help of Eqs.~(\ref{d1}) and (\ref{f1}), the
requirement~(\ref{flucnew}) becomes
\begin{equation} \label{f2}
\gamma \sim e^2 \nu (N/R)^{1/4}/\kappa >e^2/\kappa R~.
\end{equation}
Therefore only at $\nu > \nu_s \equiv 1/(NR^3)^{1/4}$ can we have
big enough potential fluctuation. At $\nu >\nu_s$ BDOGS is smeared
so strongly that at a given $\nu$ BDOGS $g(\varepsilon) \sim g_s$
at all energies, and $g_0(\mu)$ does not depend on the position of
$\mu$ any longer.

\vspace{8pt}
 { \bf  2. Line of maxima of oscillating BDOGS.}
\vspace{8pt}

\begin{figure}[ht]
\begin{center}
\includegraphics[width=0.48\textwidth]{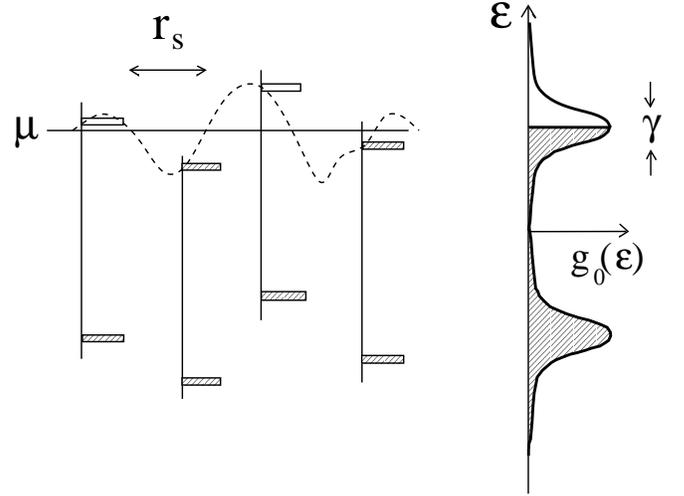}
\end{center}
\caption{ The origin of $g_{max}$. The solid horizontal line is
the Fermi level, the dashed line is the fluctuating potential
energy. Two peaks of BDOGS of the array are shown, as well as two
corresponding levels of 4 dots.} \label{fig:max}
\end{figure}

When $\nu$ is close to a half-integer $M-1/2$ ($M$ is an integer)
the Fermi level is in the middle of the $M$th peak in the BDOGS of
dots (Fig.~\ref{fig:max}).

The single dot charging levels vary with the electric potential
energy. The typical fluctuation of the potential energy is
determined by Eqs.~(\ref{d1}), which broadens the BDOGS peak
(Fig.~\ref{fig:max}). The BDOGS at the peak is
\begin{equation} \label{d2}
g_{max}(\mu) \sim g_{peak}(\varepsilon)\sim \frac{N}{\gamma}~.
\end{equation}
In such a situation the linear screening radius is
\begin{equation} \label{d3}
r_s= \frac{1}{\sqrt{4 \pi g_{max}(\mu)e^2/\kappa}}~.
\end{equation}
At $1 < \nu < \nu_s$ Eqs.~(\ref{d1}), (\ref{d2}) and (\ref{d3})
self-consistently determine all three unknowns: the BDOGS
$g_{max}(\mu)$, the screening radius and the typical fluctuation
of the potential energy. We arrive at Eq.~(\ref{bb2}) and
\begin{equation} \label{d5}
r_s\sim {\nu}^{2/3} /N^{1/3}~>>N^{-1/3}~,
\end{equation}
\begin{equation} \label{d4}
\gamma \sim e^2 {\nu}^{4/3} N^{1/3}/\kappa~.
\end{equation}

It is clear from Eq.~(\ref{bb2}) that $g_{max}$ arrives at the
value $g_s\sim N/(e^2/ \kappa R)$ when $\nu \sim \nu_s \equiv
1/(NR^3)^{1/4}>>1$. Note that the theory of this subsection is
similar to Ref.~\cite{Stern}.

\vspace{8pt}
 { \bf  3. Line of minima of oscillating BDOGS.}
\vspace{8pt}

\begin{figure}[ht]
\begin{center}
\includegraphics[width=0.48\textwidth]{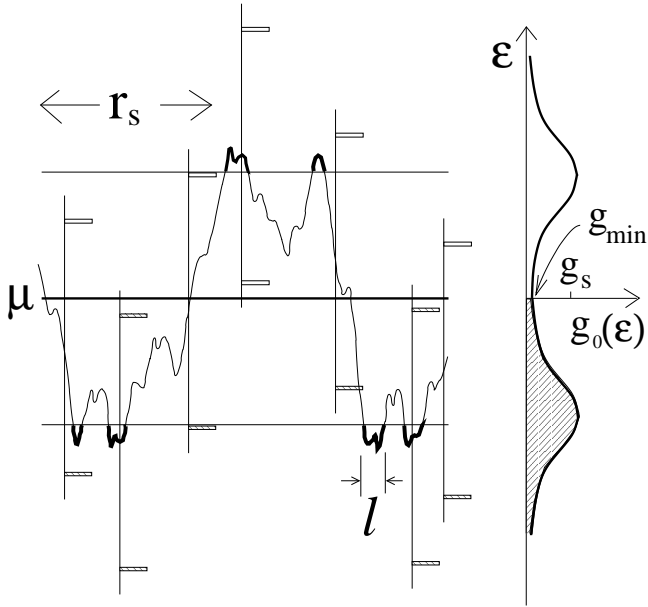}
\end{center}
\caption{The origin of $g_{min}$. The meandering line represents
the fluctuating potential energy as a function of coordinate. The
regions of the shortest size $\sim l$ where dots are completely
filled with electrons or holes are shown with thicker lines. The
solid horizontal line in the middle is the Fermi level, the other
two horizontal lines correspond to the nearest peaks of an
isolated dot. Two peaks of BDOGS of the array are shown, as well
as two corresponding levels of 7 dots.} \label{fig:min}
\end{figure}

When $\nu$ is close to an integer $M$ the Fermi level lies in the
tail between the $M$th and $(M+1)$th peaks in the BDOGS of dots.
Due to the low BDOGS at tails, screening of the long-range
fluctuations is weak and strongly nonlinear. The random potential
fluctuation grows until it reaches $e^2 /2 \kappa R$, where levels
of some dots cross the Fermi level (Fig.~\ref{fig:min}). This
picture is similar to the model of compensated semiconductor and
we treat it below following ideas of Chapter 13 of
Ref.~\cite{book}. Let us define $r_s$ as nonlinear screening
radius. Then all space can be divided into blocks of size $r_s$
with fluctuating excess charges $q \sim\nu e \sqrt{N r_s ^3}$
which create potential $\gamma(r_s) \sim \frac{\nu e^2}{\kappa
r_s}\sqrt{N r_s ^3}$. This requires
\begin{equation} \label{e2}
e^2/\kappa R \sim \gamma(r_s) \sim \frac{\nu e^2}{\kappa r_s}
\sqrt{N r_s ^3}
\end{equation}
and we arrive at
\begin{equation} \label{e1}
r_s \sim \frac{1}{N \nu ^2 R^2} ~.
\end{equation}

The potential fluctuation forms wells and humps
(Fig.~\ref{fig:min}). Electrons fall into wells and holes fall
into humps. We will discuss the wells and electrons, although our
discussion is equally applicable to humps and holes. Inside a well
of size $r_s$ there are other humps and wells of shorter range.
Electrons inside wells find themselves in wells of smaller scale
and so on until they reach the shortest scale $l$.

The shortest range $l$ can be determined by the fact that each dot
accepts only one electron. The excess number of charges inside
such a well is of the order $\nu \sqrt{Nl^3}$. The maximum number
of electrons in the well is of the same order, otherwise the well
turns into a hump. On the other hand the maximum number of
electrons is $Nl^3$ so that $Nl^3 \sim \nu \sqrt{Nl^3}$. This
gives
\begin{equation} \label{e0}
l\sim(\nu ^2/N)^{1/3}
\end{equation}
for the characteristic size of the shortest scale well, and
\begin{equation} \label{e10}
\gamma(l)\sim \frac{\nu e^2}{\kappa l} \sqrt{N l ^3} \sim \nu
^{4/3} N^{1/3} e^2/ \kappa~.
\end{equation}
for the characteristic depth. It can be checked that $l<<r_s$ as
long as $\nu <<\nu_s$, in accordance with our assumption.

As we said, in a cube of size $r_s$, the fluctuation of extra
charge is $q \sim\nu e \sqrt{N r_s ^3}$, which is balanced by the
screening charge $\Delta N r_s ^3 e$ ($\Delta N$ is concentration
of dots participating in screening):
\begin{equation} \label{e3}
\Delta N r_s ^3 e \sim  \nu e \sqrt{N r_s ^3}~,
\end{equation}
therefore
\begin{equation} \label{e5}
\Delta N \sim  \nu ^4 N^2 R^3 ~.
\end{equation}

With the help of Eqs.~(\ref{e10}) and (\ref{e5}), one can estimate
the BDOGS at the Fermi level as $g_{min}(\mu)\sim \Delta N /
\gamma(l)$, and arrive at Eq.~(\ref{bb3}). For peaks of
$g(\varepsilon)$ at $\varepsilon = \pm e^2/2 \kappa R$ we have
$g(\varepsilon) \approx g_s >> g_{min}(\mu)$ (see
Fig.~\ref{fig:min}).

At $\nu \sim \nu_s$  the shortest scale of wells and humps $l$
approaches the longest scale of wells and humps $r_s$, and each
dot can accept more than one electrons or holes because $\gamma >
e^2/\kappa R$. Therefore the non-linear screening crosses over to
linear screening at $\nu \sim \nu_s$. At the same time the
fluctuation of potential energy $\gamma$ is big enough to smear
the BDOGS. It can be checked that $g_{min}$ and $g_{max}$ merge to
$g_s$ at $\nu =\nu_s$.

\vspace{8pt} {\bf  4. Growing part of BDOGS.} \vspace{8pt}

When concentration of impurities is rather low, $\nu<<1$, an
electron released by a donor would like to stay on a dot close to
the donor. The common situation is a donor accompanied by one
charged dot (1-complex), while some donors stay alone (0-complex),
and some donors are accompanied by two charged dots (2-complex).
Charge conservation requires $N_0=N_2$, where $N_0$ and $N_2$ are
concentrations of 0- and 2-complexes.

At $\nu<<1$ nearly all donors are independent, and $N_0$, $N_1$
and $N_2$ are proportional to $\nu$. Therefore the Fermi level
remains a constant and the BDOGS grows linearly with $\nu$ at
$\nu<<1$.

The details of this problem is the same as in a weakly compensated
doped semiconductors. It has been studied
quantitatively~\cite{compensated}, counting the numbers of 0- and
2-complexes. The quantitative results are
\begin{equation} \label{n0}
N_0 = N \nu \exp \left(- \frac{4\pi}{3} \frac{e^2 N}{\kappa
|\mu|^3} \right)~,
\end{equation}
\begin{equation} \label{n2}
N_2=7.14 \cdot 10^{-4} \cdot  \left(\frac{4\pi}{3} \right)^2 \cdot
N\nu \left(\frac{e^2 N^{1/3}}{\kappa \mu} \right)^6~,
\end{equation}
\begin{equation} \label{c6}
\mu = -0.99 {e}^2{N}^{1/3}/\kappa~.
\end{equation}
As the concentration of impurities $N \nu=N_0+N_1+N_2$ is not
changed, the concentration of electrons on dots is controlled by
the Fermi level $\mu$, therefore
\begin{equation} \label{c7}
g_<(\mu)=\frac{dN_2}{d\mu}-\frac{dN_0}{d\mu}=0.2\kappa \nu
 N^{2/3}/{e}^2~.
\end{equation}
We kept all coefficients in this subsection because they are
known.

Thus $g_<(\mu)$ is proportional to $\nu$ and follows
Eq.~(\ref{bb1}). This picture works well at $\nu << 1$. At $\nu
\ge 1$ each dot is typically affected by more than one impurities.
It is straightforward to check that Eq.~(\ref{bb1}) matches
Eq.~(\ref{bb2}) at $\nu \sim 1$.

\vspace{8pt} { \bf  5. Periodic array of dots.} \vspace{8pt}

Above we have discussed sparse arrays in which the dots are
situated randomly. Now we turn to a question, what if the dots are
arranged periodically in space?

The qualitative picture in Fig.~\ref{fig:curve} is still mostly
correct. At $\nu>>1$ there are still the oscillations of
$g_0(\mu)$ with $\nu$. One should notice that the strongly charged
periodic dots do not contribute to the fluctuations of electric
potential and only donors produce the fluctuations. As we
discussed above these fluctuations are weaker than in the random
sparse array. Therefore more donors are needed to smear BDOGS to
the same extent. In other words, the lines of maxima and minima in
this case converge slower and the value of $\nu_s$ is larger.

The linear growth of BDOGS at $\nu<<1$ shown in
Fig.~\ref{fig:curve} is not valid in a periodic array, since 0-
and 2-complexes originate from the random distribution of dots in
space. In periodic array a typical donor donates an electron to
the nearest dots and forms an electric dipole. It takes additional
energy to take an electron out of a dipole and put it near another
dipole. Therefore a hard gap is formed at the vicinity of the
Fermi level. Only the rare cases where several donors are close to
each other can produce non-zero BDOGS at the Fermi level. As a
result $g_0(\mu)$ does not grow linearly with $\nu$, but with
higher power. The similar situation will be studied more
thoroughly in the dense array of Sec. V. Therefore, here we gave
only a brief description of the result.

\section{BDOGS in a sparse two-dimensional array}
\vspace{8pt}

The $g_0(\mu)$ as a function of $\nu$ in 2D looks similar to 3D
(Fig.~\ref{fig:curve}). It also has growing part, oscillating part
and saturation part. However, in 2D oscillations survive longer
than in 3D, because the collective potential in 2D is much weaker.
Indeed, the fluctuation of potential energy in 3D grows with $r_s$
as a power law
\begin{equation} \label{g2}
\gamma _{3D}\sim \frac{\nu e^2}{\kappa r_s}\sqrt{N r_s ^3}
~\propto r_s^{1/2},
\end{equation}
while in 2D the long-range fluctuation of potential energy grows
slower than logarithmically with $r_s$
\begin{equation} \label{g3}
\gamma _{2D}\sim \left(\int\limits _{1/\sqrt{N}}^{r_s} (\frac{\nu
e^2}{\kappa r})^2 Nr dr \right)^{1/2} \propto \sqrt{\ln(\sqrt{N}r_s)}.
\end{equation}

Therefore the line of minima of oscillating part in 2D is quite
different from that of 3D. The tail in the $g(\varepsilon)$ vs.
$\varepsilon$ curve is produced here by Coulomb interaction of
nearest neighbor dots, and the minimum value of $g$ appears to be
(at $1<<\nu<\nu_s=1/\sqrt{NR^2}$)
\begin{equation} \label{g1}
g_{min} \sim N^2 (\nu e^2/\kappa)^2 /\varepsilon~|_{\varepsilon
\sim e^2/(\kappa R)}\sim \nu^2 N ^2 R^3 \kappa/e^2.
\end{equation}

The 2D linear screening radius produced by $g_{min}$ is
\begin{equation} \label{g6}
r_{s} =  \frac{\kappa}{2 \pi g_{min}e^2}\sim \frac{1}{\nu ^2 N^2
R^3}.
\end{equation}
One can verify using Eq.~(\ref{g3}) that $\gamma_{2D}<<e^2/ \kappa
R$ for such $r_s$ and therefore long-range contribution can be
neglected.

The growing part at $\nu<1/2$ can be studied in the similar way as
in 3D. That is, one can use the idea about 0-complex, 1-complex
and 2-complex to find the Fermi level
\begin{equation} \label{h2}
\mu \sim -{e}^2{N}^{1/2}/ \kappa
\end{equation}
and BDOGS at the Fermi level (at $\nu << 1$)
\begin{equation} \label{b6}
g_<(\mu)\sim \kappa \nu N^{1/2}/{e}^2.
\end{equation}

For the maximum line of oscillating part, the collective typical
fluctuation of potential energy is given by Eq.~(\ref{g3}) and
BDOGS at the Fermi level is
\begin{equation} \label{h4}
g_{max}(\mu) \sim\frac{N}{\gamma} \sim\frac{N^{1/2} \kappa}{e^2
\nu}~~~~~~~~1<<\nu<\nu_s=1/\sqrt{NR^2}.
\end{equation}
Here we dropped the logarithmic factor and therefore $r_s$
disappeared from the equations. In this approximation unlike 3D
situation, we do not need the self-consistent method. As a result
$g_{max}(\mu)$ decreases with $\nu$ much slower than in 3D.

Both Eqs.~(\ref{h4}) and (\ref{g1}) reach $g_s(\mu)\sim
N/(e^2/\kappa R)$ at $\nu\sim \nu_s=1/\sqrt{NR^2}$, where the
distinction between peaks and valleys disappears.

Up to now we discussed charging of two-dimensional array of dots
by donors or a gate situated outside of dots. In some
two-dimensional systems electrons may be provided to a dot by
donors located close to the dots. This can be done, for example,
by creating a two-dimensional gas with help of close layer of
donors parallel to its plane and then by etching outside dots both
the gas and the donor layer. If such dots are essentially
two-dimensional and heavily doped, internally induced fluctuations
can cause their charging, in contrary to what we said in
introduction about three-dimensional dots. Indeed, in this case,
fluctuations of number of electrons in the dot and correspondingly
of the Fermi level decrease inverse proportionally to the dot
radius. This decay is similar to the charging energy, so that
ratio of these energies is just proportional to the ratio of
kinetic and potential energies of electron  in the
dot~\cite{Matvei}.

\section{Dense three-dimensional arrays of dots}

Until now we dealt with sparse arrays of dots where $NR^3<<1$ as
shown in Fig.~\ref{fig:x}. Let us discuss what happens when the
small parameter $NR^3$ grows. The saturation filling factor $\nu_s
=(NR^3)^{-1/4}$ decreases, which means the number of oscillations
decreases as well. The oscillations disappear at $\nu_s \sim 1$
when $NR^3 \sim 1$, i. e. when the typical separation between dots
is of the order of the dot size.

Let us now concentrate on the other extreme (Fig.~\ref{fig:dense})
where say cubic dots are closely packed in a cubic lattice with
the period $R$ and the width $d$ of an insulator between dots is
small ($d<<R$). The insulator is uniformly doped by donors. We
show below that at a large enough average number of donors per
dot, $\nu > 1$, the BDOGS is smeared. On the other hand at
$\nu<<1$ typical donors do not contribute to BDOGS at the Fermi
level but rare clusters of donors result in a small BDOGS.

For the system of metallic cubic dots \{$i$\} separated by a clean
insulator, potentials $\phi_i$ and charges $Q_i$ of dots are
related by linear equations
\begin{equation} \label{define}
\phi_i=\sum_{j}{P}_{ij}~Q_j
\end{equation}
and
\begin{equation} \label{define2}
Q_i=\sum_{j}{C}_{ij}~\phi_j~,
\end{equation}
where $P_{ij}$ is the reciprocal matrix of the capacitance matrix
$C_{ij}$. The coefficients satisfy $P_{ij}=P_{ji}$, $0 \le P_{ij}
\le P_{ii}$ and depend on the relative positions of $i$ and $j$
only. Without donors the Hamiltonian can be written as
\begin{equation} \label{ham}
H=\frac{1}{2}\sum_{ij}{P}_{ij}~Q_iQ_j=\frac{1}{2}\sum_{i}{P}_{ii}
~Q_i^2 +\frac{1}{2} \sum_{i \ne j}{P}_{ij}~Q_iQ_j~.
\end{equation}
Of course, it has minimum when all $Q_i=0$.

\begin{figure}[ht]
\begin{center}
\includegraphics[width=0.45\textwidth]{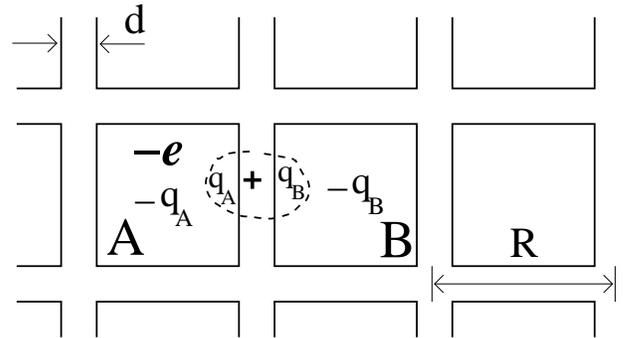}
\end{center}
\caption{ Two-dimensional cross section through a donor (with
charge $+e$) located between the dots A and B for the system shown
in Fig.~\ref{fig:dense}. The group within the dashed circle is
neutral altogether and does not affect the energy levels of the
dots A and B. The dot A gets an electron donated by the donor
because the donor is closer to its border.} \label{fig:screen}
\end{figure}

In order to understand the role of ionized donors let us study an
isolated donor between the dots A and B (Fig.~\ref{fig:screen}).
The positive donor is completely screened by negative charges
$q_A$ and $q_B$ appearing of the surface of A and B, i. e.
$q_A~+~q_B~+~e=0$. These charges leave compensating positive
charges $-q_A$ and $-q_B$ in the dots A and B. If $x_A$ and $x_B$
are distances from the donor to the dots A and B respectively then
\begin{equation} \label{qx}
q_Ax_B=q_Bx_A.
\end{equation}
One can easily arrive at Eq.~(\ref{qx}) imagining that the donor
for a moment is replaced by the uniformly charged plane parallel
to the faces of cube A and B (with the same $x_A$ and $x_B$), and
minimizing the sum of Coulomb energies of two emerging plane
capacitors with distances between planes $x_A$ and $x_B$ as a
function of $q_A$. Note that at $d<<R$ all points of the charged
plane are in the same conditions and the plane fields are just a
superposition of fields of these charges. This means
Eq.~(\ref{qx}) should also hold for a point charge of this plane
or just for a donor. Eq.~(\ref{qx}) means that potentials of dot A
and B are equal. In other words, the group made of the positive
donor, $q_A$ and $q_B$ produces an electric field which is
confined in the vicinity of the donor. The group does not affect
the electric potential of any dot. Therefore the group can be
totally forgotten when we study the charges and potentials of the
dots, or in other words it can be totally ignored in the
Hamiltonian.

On the other hand, the compensating positive charges $-q_A$ and
$-q_B$ can not be ignored. One can say that the donor's charge is
split into two fractional parts $-q_A$ and $-q_B$. Charges $-q_i$
are uniformly distributed between $0$ and $e$.

When there are more than one donor around a dot A we can simply
add them up $Q_{AD}= -q_{A1}-q_{A2}- \cdots$ to get the total
positive charge induced by donors on the dot A.

Each donor also donates an electron to the array of dots. We
assume that $d>>a$, where $a$ is the tunneling length of electrons
in the insulator so that the conductance $G$ between two dots is
much smaller than $e^2/h$. In this case each dot can get only
integer number of extra electrons $n_i$. The neutrality of the
whole system requires $-e\sum n_i=\sum Q_{iD}$. For a given
fractional set \{$Q_{iD}$\} the integer set \{$n_i$\} minimizes
the Hamiltonian
\begin{equation} \label{lowham}
H=\sum_{i}\frac{{P}_{ii}}{2}(Q_{iD}-n_ie)^2+ \sum_{i \ne
j}\frac{{P}_{ij}}{2}(Q_{iD}-n_ie)(Q_{jD}-n_je).
\end{equation}
This energy is calculated from the reference point of the sum of
energies of neutral groups around each donor, which does not
depend on {$n_i$}. If each donated electrons were shared by dots
this fractional charges would neutralize {$Q_{iD}$} and make $H=0$
and all $\phi_i = 0$. This is equivalent to what would happen if
all grain are grounded. Discreteness of {$n_i$} makes finding the
ground state {$n_i$} nontrivial. It is clear that in the globally
neutral system the chemical potential $\mu$ lies in the same place
as for a system of neutral dots, i. e. $\mu=0$. In the ground
state we can introduce ``one-dot" energies of an empty dot state
above the Fermi level assuming that charges of other dots are
fixed
\begin{equation} \label{up}
\varepsilon_i^{(e)}=H(n_i+1)-H(n_i)=-e\phi_i+\frac{1}{2}P_{ii}e^2~,
\end{equation}
where
\begin{equation} \label{pot}
\phi_i=\sum_{j}{P}_{ij}~(Q_j-n_je)
\end{equation}
is the electrostatic potential of the dot $i$. For an occupied
state in the similar way
\begin{equation} \label{dn}
\varepsilon_i^{(o)}=H(n_i)-H(n_i-1)=-e\phi_i-\frac{1}{2}P_{ii}e^2~.
\end{equation}
In the absence of donors when all dots are neutral, i. e.
$\phi_i=0$, we get $\varepsilon_i^{(e)}=+\frac{1}{2}P_{ii}e^2$ and
$\varepsilon_i^{(o)}=-\frac{1}{2}P_{ii}e^2$. The single electron
BDOGS for neutral dots looks the same as in Fig.~\ref{fig:init}
except the separation of peaks or the level spacing is smaller now
$e^2P_{ii}\sim e^2/\kappa_{eff} R \simeq \frac{e^2d}{ \kappa R^2}$
(the calculation of $P_{ii}$ is in Appendix). Here
\begin{equation} \label{dielectric}
\kappa_{eff} \simeq \kappa R/d
\end{equation}
is the effective dielectric constant of the array. For charged
dots stability of the global ground state requires
$\varepsilon_i^{(e)}>\mu=0$ and $\varepsilon_i^{(o)}<\mu=0$.

Let us study the BDOGS in the case when there are more than one
donors per dot ($\nu>>1$). In order to find BDOGS we neglect the
interaction terms in Eq.~(\ref{lowham}) and consider only diagonal
terms. As a result the ground state of the system is realized by
such a set of \{$n_i$\} that $-e/2 \le Q_{iD}-n_ie \le e/2$. So
effective charges on different dots in the ground state of the
array are uniformly distributed from $-e/2$ to $e/2$. This gives
the energy independent BDOGS
\begin{equation} \label{sme}
\tilde{g}_s(\varepsilon)\sim \frac{1}{e^2 P_{ii}R^{3}}~~ .
\end{equation}
Consider now the role of non-diagonal terms. At a big distance
$r_{ij}>>R$ the coefficients $P_{ij}$ have the Coulomb form
$P_{ij} \approx 1/ \kappa_{eff} r_{ij}$. They, therefore, result
in a density of ground states (DOGS) with the Coulomb gap at the
Fermi level, which in turn leads to ES law.

The finite BDOGS at the Fermi level and ES law at $\nu>>1$,
originates from random distribution of charges $Q_i$, which in
turn is related to the fact that several donors contribute into
each $Q_{iD}$. Even for nearest neighbor dots $Q_{iD}$ and
$Q_{jD}$ are independent since they include contributions of
different donors.

\begin{figure}[ht]
\begin{center}
\includegraphics[width=0.21\textwidth]{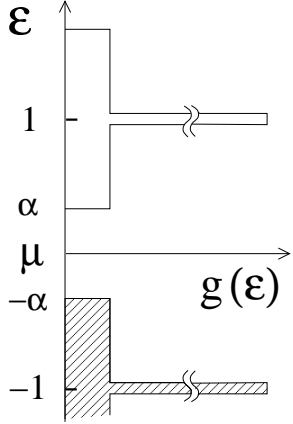}
\end{center}
\caption{BDOGS of isolated donors at $\nu<<1$. Energy
$\varepsilon$ is measured in units of $\alpha P_{ii}e^2/2$. There
is a hard gap of the width $\alpha P_{ii}e^2$ around the Fermi
level. } \label{fig:hard}
\end{figure}

We turn now to the case of a small density of donors $\nu<<1$ and
first deal with typical donors isolated from each other. As
discussed above for such an isolated donor between the dots A and
B, the charges induced by the donor are related by
$Q_{AD}+Q_{BD}=+e$. If the donor is closer to the dot A,
$Q_{AD}>e/2>Q_{BD}>0$. If we neglect interaction between the dots
A and B in Eqs.~(\ref{up}) and (\ref{dn}) we arrive at finite
BDOGS of isolated donors at the Fermi level. States at the Fermi
level are provided by dots where $Q_{BD}=1/2-\delta$ with
$\delta<<1$. Then the dot B has a state just above the Fermi level
and A has a state just below it. The fact that these states are so
closely correlated in space leads to their elimination by the
interaction between the dots A and B. Indeed, in the ground state
the electron donated by the donor is on the dot A. The total
charges on A and B is $Q_A=Q_{AD}-e=-Q_{BD}$ and $Q_B=Q_{BD}$. The
electric potential on B is $\phi_B=P_{BB}Q_B+P_{BA}Q_A$ and
according to Eq.~(\ref{up}) all the levels on B are moved down by
$e\phi_B$. Using the fact $0 \le Q_{BD} \le e/2$, we have
\begin{equation} \label{down}
0 \le e\phi_B \le (1-\alpha)P_{BB}~e^2/2~,
\end{equation}
where $\alpha \equiv P_{BA}/P_{AA}=0.34 \pm 0.01$ represents the
interaction between nearest neighbor dots (see the calculation of
$\alpha$ in Appendix). According to Eq.~(\ref{up}) the lowest
vacant level of dot B is at least $\alpha P_{ii} e^2/2$ above the
Fermi level. Similarly, according to Eq.~(\ref{dn}) the highest
filled level on the dot A is moved up by $-e\phi_A$ with $0 \le
-e\phi_A \le (1-\alpha)P_{AA}e^2/2$ and it is at least $\alpha
P_{ii} e^2/2$ below the Fermi level (Fig.~\ref{fig:hard}).

Thus, we showed that for $\nu<<1$ typical isolated donors have a
substantial hard gap of width $\alpha P_{ii}e^2$ in their density
of states. This gap can not be destroyed by long range
interactions of such donors because they form neutral complexes
and weakly interact with each other at long distances.

One can also think of a donor in the neutral complex with electron
of the neighboring dot A as a hydrogen-like donor in a
semiconductor. Then the hard gap obtained above is similar to
Hubbard gap. The electric dipole made of a donor and an electron
in principle can accept another electron with a weak binding
energy, but the energy difference or Hubbard gap between the two
electronic levels makes sure that only the lower level is filled
in the ground state.

Up to now we talked about typical isolated donors at $\nu<<1$. In
fact there are rare configurations that make a non-zero BDOGS at
the Fermi level. Let us show how this happens.

\begin{figure}[ht]
\begin{center}
\includegraphics[width=0.25\textwidth]{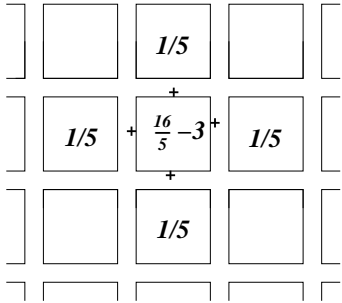}
\end{center}
\caption{ An example of rare configurations leading to the finite
BDOGS at the Fermi level at $\nu<<1$. The cluster of 5 dots and 4
donors is effectively positively charged in the ground state.
Charges of 5 dots are shown in units of $e$. All the other dots in
the vicinity of the cluster remain neutral.} \label{fig:cross}
\end{figure}

Consider, for example, a cluster of 4 donors around one dot
(Fig.~\ref{fig:cross}). The charges of donors are shared by 5 dots
in such a way that charge of each donor is effectively split into
$4e/5$ going to the central dot and $e/5$ going to its nearest
neighbor. The middle dot has effective charge $Q_{iD}=16e/5$ and
each of the four neighbor dots has $Q_{jD}=e/5$. Let us show that
in the ground state of the cluster $n_i=3$ for the central dot and
$n_i=0$ for all others, i. e. there are only three electrons on
the cluster of four donors so that the cluster has a net charge
$+e$. It is effectively shared by the 5 dots, each getting $+e/5$.
According to Eq.~(\ref{up}) addition of the fourth electron to
this cluster costs
\begin{equation} \label{cost}
\varepsilon_i^{(e)}=\left[-\frac{1}{5}(4\alpha+1)+\frac{1}{2}
\right]P_{ii}e^2 \approx 0.03 P_{ii}e^2
\end{equation}
where $\alpha=P_{12}/P_{11}$, and as shown in Appendix
$\alpha=0.3405$. This means that $\varepsilon_i^{(e)}>\mu =0$ and
the positively charged cluster of Fig.~\ref{fig:cross} is stable
with respect of bringing the fourth neutralizing electron. The
physical reason for such stability is uniform smearing of the net
charge over 5 dots which diminishes potential of the cluster
$\phi_{i}$ at the central dot attracting the fourth electron (the
first term in the right side of Eq.~(\ref{cost})).

We have shown above an example of a positive ``donor-like"
cluster. Moving each of the four donors away from the central dot
in the direction of the neighbor dot one can construct an
``acceptor-like" cluster in which charge $-e$ is evenly shared by
5 dots. Such a cluster accepts an electron released by the cluster
in Fig.~\ref{fig:cross}. Thus, some clusters charge themselves by
``self-compensation" similarly to amphoteric impurities in
semiconductors. Of course, there are also neutral clusters and
tuning positions of donors between dots one can continuously go
from a neutral cluster to the charged one. This means that BDOGS
at the Fermi level created by clusters is finite.

Using coefficients $P_{ij}$ calculated in Appendix, it is easy to
find that the smallest charged cluster in the ground state is the
cluster of 4 dots with 3 donors between them. Therefore, at $\nu
<<1$ the BDOGS at the Fermi level grows as $g_0(\mu) \propto
\nu^3$ and it begins to be significant at $\nu \sim 1$ (see
Fig.~\ref{fig:curvenew}).

The structure of self charging clusters discussed above is similar
to the mechanism which provides a non-zero bare density of states
in a system of quasi-one-dimensional electron crystals at low
impurity concentration~\cite{Fogler}. There is also similarity to
self-compensation of clusters of several donors in uncompensated
semiconductor near the insulator-metal transition~\cite{Bhatt}.

As we saw above in the ``super-dense" array of dots with $d<<R$, a
large enough concentration of donors can easily smear BDOGS. This
is a natural explanation of the experiments, where the ES law is
observed. Let us consider parameter $T_0$ of the ES law
(Eq.~(\ref{a1})) for a dense array of dots at $\nu>>1$. According
to Ref.~\cite{book} $T_0\sim {e^2}/(\kappa_{eff} \xi)$, where
$\xi$ is the localization length for tunneling to distances much
larger than $R$. When an electron tunnels through insulator it
accumulates dimensionless action $d/a$, where $a$ is the tunneling
decay length in the insulator. On the distance $x$ electron
accumulates $x/R$ such actions. Thus, its wave function decays as
$\exp(-{x}{d}/ {R}{a}) = \exp(-x/\xi)$, where the localization
length $\xi =aR/d$. This enhancement of localization length is
similar to the one derived for disordered granular system near
percolation threshold~\cite{Sh}. It leads to
\begin{equation}
\label{tempe} T_0\sim\frac{e^2d}{\kappa_{eff}
aR}=\frac{e^2d^2}{\kappa R^2a}~.
\end{equation}
We see that with decreasing $d$ the temperature $T_0$ decreases as
$d^2$. This continues until conductance of the insulator layer
$G=(e^2/\hbar)(Rk_F)^2 e^{-2d/a}$ reaches the quantum limit $G\sim
e^2/\hbar$, i. e. while $d>d_c \equiv a \ln(Rk_F)$. At $d<<d_c$
the characteristic ES temperature $T_0(d)$ should vanish when
$d\rightarrow 0$, but the way how this happens is still unknown
(see also recent publications~\cite{Turlakov,Glazman} concerned
with the case $G>>1$).

In the above calculation of the localization length we assumed
that tunneling through a metallic dot does not accumulate more
action. Actually a dirty dot provides logarithmic contribution to
the tunneling action~\cite{Averin}. We assume that it is much
smaller than $2d/a$ resulting from crossing the insulator.

\begin{figure}[ht]
\begin{center}
\includegraphics[width=0.25\textwidth]{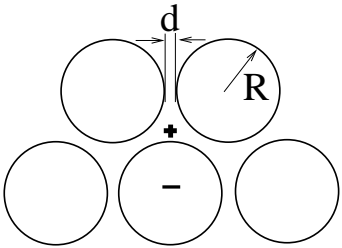}
\end{center}
\caption{An almost densely packed array of metallic spheres
($d<<R$). A donor is shown by + and electron donated by it is
shown by -. } \label{fig:almost}
\end{figure}

Until now we discussed only the ``super-dense" array shown in
Fig.~\ref{fig:dense}. Let us consider what happens in the system
of almost densely packed metallic spheres (Fig.~\ref{fig:almost}),
where insulator occupies a larger fraction of space than in
Fig.~\ref{fig:dense}.

One can show that in such a system BDOGS as a function of number
of donors per dot $\nu$ behaves qualitatively similar to
Fig.~\ref{fig:curvenew}. At smaller $\nu$ only a cluster of donors
around a dot can create a finite BDOGS, while at $\nu>>1$ donors
easily smear BDOGS. In this case, characteristic temperature of ES
law is $T_0 \approx e^2d/aR$. It loses one power of $d/R$ because
the dielectric constant of such systems does not diverge at
$d\rightarrow 0$. When distance between dots $d$ grows and exceeds
$R$, we get sparse periodic array and recover oscillations of
BDOGS at $\nu>>1$. At $\nu<<1$ there is a smooth cross-over
between two super-linear growths of BDOGS related to rare donor
clusters in the vicinity of some dots.

Although we arrived at $\nu>>1$ as the universal criterion of
substantial smearing of BDOGS for both dense arrays
(Figs.~\ref{fig:dense} and \ref{fig:almost}) and sparse
arrays(Figs.~\ref{fig:x}), we should understand that the required
concentration of donors $N_D$ in the insulator grows while volume
fraction of insulator decreases from the arrays on
Figs.~\ref{fig:x} and to the one on Fig.~\ref{fig:almost} and then
to the one on Fig.~\ref{fig:dense}.

\begin{figure}[ht]
\begin{center}
\includegraphics[width=0.25\textwidth]{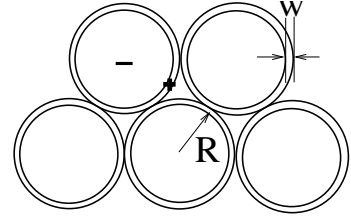}
\end{center}
\caption{An array of coated dots. The width of coating insulator
is $w$. A donor is shown by + and electron donated by it is shown
by -. } \label{fig:coat}
\end{figure}

In some cases metallic granules are first coated by the layer of a
doped insulator (for example, the metal's oxide) and then
compressed into the bulk array with a clean insulator of different
kind of filling or just air filling empty space
(Fig.~\ref{fig:coat}). For simplicity we assume the dielectric
constant of the coating insulator is close to that of the filling
space between dots.

We would like to consider BDOGS for such a model of granular metal
in order to understand cross-over between the studied above case
of donors residing outside dots and those on the dot surface or
inside dots. In the latter case, donors are screened and act as
short range neutral impurities. According to the introduction such
impurities can not charge a neutral array of dots. Thus, effect of
donors in the layer of the width $w << R$ should decrease as $w$
vanishes.

Indeed, one can calculate, fluctuation of average the potential of
donor layer of the width $w$ which plays the role of fluctuation
of work function of dots. It can be estimated as a potential
created by the layer of the width $w$  by fluctuating
concentration $(N_D wR^{2})^{1/2}/wR^{2}$. One can imagine that,
correction to the work function is created by effective random
double layer. Such correction to the work function is of order
$e^2((N_D w^{3})^{1/2} /\kappa R$. It becomes larger than the
charging energy only at $N_D w^{3} >>1$~\cite{secondmethod}. This
condition of BDOGS smearing is obviously much stronger than ``one
donor per dot" condition valid for a uniform insulator. When $w$
approaches the lattice constant the role of donors is so much
weakened that in order to smear BDOGS one needs the relative
concentration of donors to be of the order of $50\%$. At the
surface, donors and their images are still forming dipoles with
fluctuating dipole moments, which create fluctuating double layer
potential. On the other hand, when donors enter inside the dot,
each donor becomes exponentially screened from all sides, the
dipoles and the random double layer disappear. This diminishes the
role of donors further so that as we said in introduction even
large concentration of impurities inside the whole three
dimensional dot can not charge the dot.

\section{Conclusion}

This paper addresses the problem of explanation of the temperature
dependence of the conductivity observed in granular metals and
arrays of quantum dots. We show that ES law is the most natural
explanation for this dependence (in spite of recently expressed
doubts~\cite{Efetov}). For this purpose we present a simple model
with random charging of clean metallic dots in both sparse and
dense arrays of dots. In both cases we assume that the insulator
separating dots is uniformly doped by donors, which donate their
electrons to dots. Random positions of dots in the sparse case
(Fig. 1) and random positions of donors in the dense periodic
model (Fig. 5) lead to random charging of dots in the global
ground state and to filling of the gap of the bare density of
individual dots at the Fermi level. The long range Coulomb
interaction creates a soft Coulomb gap on the background of the
finite bare density of ground states in the vicinity of the Fermi
level.  At low enough temperature this leads to the ES variable
range hopping conductivity, in agreement with multiple
experimental observations.

We concentrate on the dependence of the bare density of states on
a number of donors per dot. Such dependence for the sparse and
dense models are shown in Fig. (7) and Fig. (9) are qualitatively
different. The sparse random model shows linear growth of BDOGS at
weak doping and oscillations of the density of states at strong
doping. On the other hand the dense model shows very small density
of states at weak doping and no oscillations at stronger doping.
The bare density of states determines the width of the Coulomb gap
and the characteristic temperature of transition from ES law to
Mott's law with increasing temperature. It determines many
thermodynamic properties of the array as well.

For dense arrays we have calculated characteristic temperature
$T_0$ of ES law in the case, when tunneling conductance between
granules is small. The challenging question of calculation of
$T_0$ in the case of even closer dots will be addressed in the
next publication. Another interesting question left beyond the
scope of this paper is the origin of ES law in isotropic and
anisotropic arrays of strongly  anisotropic (elongated) granules,
for example, nano-tubes.

In this paper we concentrated on ohmic hopping conductivity.
However, our results can be applied to non-ohmic conductivity,
too. It was shown~\cite{Strong} that at low temperatures a strong
electric field $E$ replaces temperature $T$ in the exponential
temperature dependence of the variable range hopping by the
effective temperature $T_E = eE\xi/2k_B$. As a result non-ohmic ES
law reads $j=j_0\exp[-(E_{0}/E)^{1/2}]$ with $E_{0} =  2
T_0/e\xi$. This dependence was observed in granular
 metals and nanocrystal thin films~\cite{sheng,Guyot}.

\begin{acknowledgments}
We are grateful to M. V. Entin, M. M. Fogler, Yu. M. Galperin, A.
Kamenev and A. I. Larkin for helpful discussions. This work was
partially supported by NSF No. DMR-9985785.
\end{acknowledgments}

\vspace{30pt} {\bf  APPENDIX: MATRIX ELEMENTS $P_{ij}$ FOR
NEIGHBORING DOTS.} \vspace{24pt}

Here we study the coefficients $P_{ij}$ in Eq.~(\ref{define}). We
know the diagonal term ${P}_{ii} \sim 1/ \kappa_{eff} R$, and for
distant neighbors $r_{ij}>>R$ one can show that ${P}_{ij}\sim 1/
\kappa_{eff} r_{ij}$. But the coefficients between close neighbors
require further consideration.

The dense array of periodic dots is equivalent to the lattice of
identical capacitors connecting adjacent dots. Such a system can
be studied in the language of an equivalent resistor network.
Imagine a cubic lattice network with an identical resistance
between every nearest neighbor sites. Suppose that current $I$
goes into a lattice site A and travels out through the network to
infinity. Label the potential of site A to be $V$, the potential
of nearby dots B, C, D, $\cdots$ to be $\alpha_{AB} V$,
$\alpha_{AC} V$, $\alpha_{AD} V$, $\cdots$. What are the values of
the $\alpha_{AB}$, $\alpha_{AC}$, $\alpha_{AC}$, and so on? Here
we are using the dimensionless coefficients such as $\alpha_{AB}
\equiv {P}_{AB}/{P}_{AA}$, and we know all of them are between $0$
and $1$.

A standard numerical calculation uses the fact that the current
going into a site equals the current going out of it (Kirchhoff
law), which means the potential of a site equals the average
potential of its nearest neighbors. Using lattices of $N \times N
\times N$ sites and keeping the potential at central cite to be
$1$ and that of boundaries to be $0$, one adjusts the potential
values of the other sites so that the potential of each site is
very close to the average potential of its nearest neighbors.
Using lattices with $N$ up to $97$ we find $\alpha \equiv
\alpha_{01}=0.34 \pm 0.005$, $\alpha_{02}=0.166 \pm 0.005$,
$\alpha_{03}=0.106 \pm 0.005$, $\alpha_{11}=0.216 \pm 0.005$,
$\alpha_{111}=0.17 \pm 0.005$ (here $111$ represents the vector
between the two dots measured in the units of the lattice
constant).

There is also an analytic way to find the
coefficients~\cite{Foglerway}. Imagine the situation when there is
charge $Q_0$ on the central dot while all the other dots remain
neutral, one can use the equation similar to Kirchhoff law
mentioned above
\begin{eqnarray} \label{kirchhoff}
\phi(\vec{r})- \frac16 \sum_{r'}\phi(\vec{r'}) =
\frac{Q_0d}{6\kappa R^2 } ~\delta_{\vec{0}, \vec{r}}~~,
\end{eqnarray}
where $\vec{r'}$ are 6 nearest neighbor of the site $\vec{r}$, and
$\phi(\vec{r})$ is defined only on the discrete sites
$\vec{r}=(n_xR,n_yR,n_zR)$. Eq.~(\ref{kirchhoff}) can be solved
using discrete Fourier transform
\begin{equation}  \label{kspace}
\phi_{\vec{k}}-\frac16\phi_{\vec{k}}\cdot ( \cos k_xR + \cos k_yR
+ \cos k_zR ) =\frac{Q_0d}{6\kappa R^2 } ~.
\end{equation}
One can find $\phi_{\vec{k}}$ and
\begin{eqnarray}  \label{findpij}
\phi(\vec{r})= \frac{dQ_0}{2\kappa R^2 } \int\limits_{BZ}
\frac{d^3 k}{(2 \pi)^3} \frac{R^3 \cos(\vec{k} \cdot \vec{r})}{3 -
\cos k_xR - \cos k_yR - \cos k_zR }
\end{eqnarray}
where the integration is over the Brillouin zone $k_x, k_y, k_z
\in (-\pi/R,\pi/R)$. One can use numerical integration to find
\begin{eqnarray}  \label{resultsp}
P_{00} \equiv\frac{\phi(\vec{0})}{Q_0}= \frac{0.2527d}{\kappa R^2
},~\alpha \equiv \alpha_{01} =
\frac{P_{01}}{P_{00}}=0.3405,\nonumber\\\alpha_{02}=0.1697~,
~\alpha_{03}=0.1089~,~\alpha_{11}=0.2183~~~~~~
\end{eqnarray}
and so on. These results agree with the results of the first
method.

\end{multicols}

\begin{thebibliography}{100}
\bibitem{criticismBaskin} Baskin and Entin~\cite{Baskin}
postulated that excited states bring some dots into infinite
percolation cluster of resistances and lead to exponentially
larger hopping conductivity than ground states. In this way they
obtained a new exponential factor of hopping conductivity where
power 2/5 replaces power 1/2 in ES law. We disagree with this
claim because filling of a dot by electrons is controlled by the
ground state.
\bibitem{secondmethod} Alternatively, one can derive the same inequality
using idea of fractionalization of a donor charge. In the case of
coated dots, the host dot of the donor gets all but a fraction of
the order of $w/R$ of its charge, which is acquired by the host
dot nearest neighbors. The non-integer part of the host dot charge
reaches $e/2$ only when fluctuation of number of donors in the
coating layer
 $(N_{D}R^{2}w)^{1/2}$
multiplied by $w/R$ becomes of the order of unity. This brings us
back to $N_{D}w^{3} \gg 1$.
\end{thebibliography}
\end{document}